\documentclass[a4paper,10pt,aps,pre,notitlepage,reprint]{revtex4-1}
\usepackage{bm,amssymb,amsmath,amsfonts,graphicx}

\newcommand{\ca}{\mathcal{A}}
\newcommand{\cb}{\mathcal{B}}
\newcommand{\cc}{\mathcal{C}}
\newcommand{\ch}{\mathcal{H}}
\newcommand{\co}{\mathcal{O}}
\newcommand{\cs}{\mathcal{S}}

\newcommand{\br}{\bm{r}}
\newcommand{\bl}{\bm{\ell}}
\newcommand{\bff}{\bm{f}}
\newcommand{\be}{\bm{e}}
\newcommand{\bT}{\bm{T}}
\newcommand{\bdr}{\bm{\delta r}}
\newcommand{\bdl}{\bm{\delta \ell}}
\newcommand{\bq}{\bm{q}}
\newcommand{\bw}{\bm{w}}

\begin{document}

\title{Forces exerted by a correlated fluid on embedded inclusions}
\author{Anne-Florence Bitbol and Jean-Baptiste Fournier}
\affiliation{Laboratoire Mati\`ere et Syst\`emes Complexes (MSC), Universit\'e Paris Diderot, Paris 7 and UMR CNRS 7057, 10 rue Alice Domon et L\'eonie Duquet, F-75205 Paris Cedex 13, France}
\date{\today}

\begin{abstract}
We investigate the forces exerted on embedded inclusions by a fluid medium with long-range correlations, described by an effective scalar field theory. Such forces are the basis for the medium-mediated Casimir-like force. To study these forces beyond thermal average, it is necessary to define them in each microstate of the medium. 
Two different definitions of these forces are currently used in the literature. We study the assumptions underlying them. We show that only the definition that uses the stress tensor of the medium gives the sought-after force exerted by the medium on an embedded inclusion. If a second inclusion is embedded in the medium, the thermal average of this force gives the usual Casimir-like force between the two inclusions.
The other definition can be used in the different physical case of an object that interacts with the medium without being embedded in it. 
We show in a simple example that the two definitions yield different results for the variance of the Casimir-like force.
\end{abstract}

\maketitle

\section{Introduction and summary}
The present work deals with the force exerted by a fluid medium with long-range correlations, described by an effective scalar field theory, on embedded inclusions. 
As inclusions constrain the fluctuations of the correlated fluid medium, the force exerted by the medium on one inclusion depends on the position of the other inclusions. Thus, medium-mediated interactions appear between inclusions. These fluctuation-induced interactions are analogous to the Casimir force, which arises between two uncharged metallic plates in a vacuum because of the boundary conditions imposed by the plates on the quantum fluctuations of the electromagnetic field \cite{Casimir48}. Casimir-like effects driven by thermal fluctuations of fluid media were first discussed by Fisher and de Gennes in the context of critical mixtures \cite{Fisher78}, and they have recently been measured \cite{Hertlein08}. Such forces appear in many other systems, including liquid crystals, fluid membranes, fluid interfaces and superfluids (for a review, see Refs.~\cite{Kardar99, Gambassi09}). 

The Casimir-like force between two inclusions is usually defined as $-\partial F/\partial \ell$, where $F$ is the free energy of the fluid medium with two inclusions separated by a distance $\ell$. However, this definition only provides a thermal average force. In order to study the fluctuations of the force, as well as out-of-equilibrium situations, it is necessary to define the force exerted on an inclusion by a fluid medium in each microstate of this medium. Two different definitions of this force are currently used: the first one uses the stress tensor of the medium~\cite{Bartolo02, Bartolo03, Najafi04, Gambassi06, Gambassi08, Bitbol10, Rodriguez11}, while the second one is based on differentiating the effective Hamiltonian with respect to the position of the inclusion while keeping constant the field that describes the state of the medium \cite{Dean09, Dean10, Demery10a, Demery10b}. The aim of the present work is to clarify the difference between these two definitions and to determine their respective domains of validity.

In this paper, we consider a correlated fluid medium described by a scalar field $\phi$: each microstate of this effective field theory corresponds to a given $\phi$. We study the force $\bff$ exerted by this fluid medium on an embedded inclusion. This force $\bff$ is defined as the negative gradient of the effective Hamiltonian $H$ of the medium together with the inclusion, with respect to the position of the inclusion. The validity of this fundamental definition in our coarse-grained description is justified in Appendix~\ref{sec_def}. In order to determine $\bff$ in a given microstate of the medium, \emph{the gradient of $H$ must be taken in this microstate}. There are two distinct ways of interpreting these words, yielding two different routes to calculate the force. In the first route, the field $\phi$ is kept constant in the Eulerian sense, i.e., $\phi$ remains the same at each point in space. In the second route, $\phi$ is kept constant in the Lagrangian sense, i.e., each fluid particle of the medium keeps the same value of $\phi$ during the infinitesimal deformation that is associated with the displacement of the inclusion. We show that the second route gives the integral of the stress tensor of the medium on the boundary of the inclusion, which corresponds to the definition used in Refs.~\cite{Bartolo02, Bartolo03, Najafi04, Gambassi06, Gambassi08, Bitbol10, Rodriguez11}. We argue that this definition is the right one for an embedded inclusion. We also consider the case of \emph{non-embedded influencing objects}, which interact with the medium without being embedded in it, and we argue that the first route, which corresponds to the definition used in Refs.~\cite{Dean09, Dean10, Demery10a, Demery10b}, is the right one in such situations.

This paper is organized as follows: in Sec.~II, we study the force $\bff$ exerted on an inclusion by the fluid medium in a given microstate, starting from the variation of the total energy of the system when it undergoes a generic infinitesimal deformation, and we compare the two routes introduced above. In Sec.~III, we proceed similarly in the case of a non-embedded influencing object. In Sec.~IV, we show the link between the thermal average of the force $\bff$ and the Casimir-like force. In Sec.~V, we study a simple example of Casimir-like force, where we show that the variance of the force depends strongly on the route that is chosen. This example illustrates the importance of choosing the right definition when studying Casimir-like forces beyond their average value at thermal equilibrium. Finally, Sec.~VI is a conclusion.

\section{Embedded inclusion}
\label{ei}
\begin{figure}[h t b]
\centerline{\includegraphics[width=6cm]{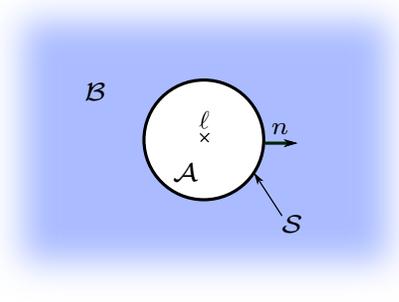}}
\caption{(Color online) Schematic representation of an inclusion (white) embedded in a two-dimensional fluid medium (shaded). The inclusion extends over $\ca$, which is centered on $\bl$, so $\phi$ is defined only on $\cb=\mathbb{R}^2\smallsetminus\ca$. The contour of $\ca$ is called $\cs$, and $\bm{n}$ denotes its outward normal.}
\label{emb}
\end{figure}

Let us consider an infinite $d$-dimensional fluid medium ($d\geq 1$) with short-range interactions, described in a coarse-grained fashion by a scalar field $\phi$ with Hamiltonian density $\ch(\phi,\bm{\nabla}\phi)$. Let us assume that an embedded inclusion with center of mass in $\bl\in\mathbb{R}^d$ extends over the region $\ca$ of this medium (see Fig.~\ref{emb}). The inclusion is made of a different material, typically a solid, where the physical field $\phi$ is not defined. 
For instance, in critical binary mixtures, $\phi$ is the order parameter, i.e., the shifted concentration in one component: $\phi=c-c^\mathrm{crit}$, where $c^\mathrm{crit}$ is the critical concentration of this component \cite{Gambassi08}, so $\phi$ is not defined in solid objects immersed in the mixture. 

Since the embedded inclusion extends over the region $\ca$, delimited by the hypersurface $\cs$, $\phi$ is defined only in the region $\cb=\mathbb{R}^d\smallsetminus\ca$ (see Fig.~\ref{emb}). We assume that the field $\phi$ is coupled to the inclusion at the interface $\cs$ through short-range interactions modeled by a potential $V(\phi)$, so the effective Hamiltonian of the system reads
\begin{equation}
H=\int_{\cb}\ch(\phi,\bm{\nabla}\phi)\,d^d r+\int_{\cs}V(\phi)\,d^{d-1} r+E_{\mathrm{incl}}\,,
\end{equation}
where $E_{\mathrm{incl}}$ represents the internal energy of the inclusion, which will not be varied in the following. 
Note that our work can be generalized to a more general coupling potential, such as $V(\phi,\bm{\nabla}\phi)$. While our proofs will be presented with $V(\phi)$ for the sake of simplicity, the final results for the force with $V(\phi,\bm{\nabla}\phi)$ will be mentioned.

Our aim is to calculate the force $\bff$, which is defined as
\begin{equation}
\bff=-\frac{\partial{H}}{\partial\ell_i}\,\be_i\,,
\label{def_force}
\end{equation}
where $\be_i$ is a unit vector in the $i$ direction. The Einstein summation convention has been used in Eq.~(\ref{def_force}), and will be used throughout. The validity of the fundamental definition Eq.~(\ref{def_force}) in our coarse-grained description is justified in Appendix~\ref{sec_def} from the principle of virtual work.

In order to obtain $\bff$, we must calculate the variation of $H$ when the inclusion undergoes the infinitesimal displacement $\bdl$. Since it is impossible to move an embedded inclusion without moving the fluid particles of the surrounding medium, we will consider a displacement field $\bdr$ in the whole medium. We use the expression ``fluid particle'' in a similar way as in fluid mechanics \cite{Batchelor}: it denotes a (macroscopically) closed mesoscopic part of the fluid medium, the state of which is described by the value of the coarse-grained field $\phi$. Let us consider the generic infinitesimal transformation
\begin{equation} 
\left\{ \begin{array}{lll}
\br & \rightarrow & \br+\bdr(\br)\,,\\
\phi(\br) & \rightarrow & \phi(\br)+\delta\phi(\br)\,.
\end{array} \right.
\label{trf_inf}
\end{equation}
More explicitly, the particle initially in $\br$ undergoes the infinitesimal displacement $\bdr (\br)$, and the value of $\phi$ at a fixed point $\br$ in space is modified by the quantity $\delta\phi(\br)$. The functions $\bdr$ and $\delta \phi$, defined respectively on $\mathbb{R}^d$ and on $\cb$, are assumed to be regular and to take small values, of order $\epsilon$ (i.e., $\delta\phi=\co(\epsilon)$ and $|\bdr|/L=\co(\epsilon)$ where $L$ is the characteristic length of interest, e.g., the distance between two inclusions in a study of the Casimir-like force). During this transformation, the total (Lagrangian) variation of $\phi$ for the fluid particle initially in $\br$ is $\delta^T\!\phi(\br)=(\phi+\delta\phi)(\br+\bdr)-\phi(\br)=\delta\phi(\br)+\bm{\nabla}\phi(\br)\cdot\bdr(\br)$ at first order in $\epsilon$. 

Let us assume that in the region $\ca$, $\bdr$ is constant, equal to $\bdl$: thus, the inclusion undergoes an infinitesimal global translation that does not affect its internal energy $E_{\mathrm{incl}}$. The total variation of the energy then reads, at first order in $\epsilon$:
\begin{align}
\delta H&=\int_\cb\left[\frac{\partial\ch}{\partial\phi}-\partial_i\left(\frac{\partial\ch}{\partial(\partial_i\phi)}\right)\right]\delta\phi\,d^{d} r\nonumber\\
&-\int_\cs\left[\ch\delta\ell_i+\frac{\partial\ch}{\partial(\partial_i\phi)}\delta\phi\right]n_i\,d^{d-1} r\nonumber\\
&+\int_\cs\frac{\partial V}{\partial \phi}\left(\delta\phi+\partial_i \phi\,\delta\ell_i\right)\,d^{d-1} r\,,
\label{vartot}
\end{align}
where we have introduced $\bm{n}$, the exterior normal to $\cs$ (see Fig.~\ref{emb}), and we have used the notation $\partial_i\phi\equiv\partial\phi/\partial r_i$ for $i\in\{1,\dots,d\}$. In this equation, the hypervolume integral on $\cb$ contains the standard Euler-Lagrange term. Meanwhile, the first term of the first hypersurface integral comes from 
\begin{equation}
\int_{\delta\cb}\ch\,d^d r=-\int_\cs \ch \delta\ell_i n_i\,d^{d-1}r\,,
\end{equation}
while its second term is obtained via Stokes' theorem:
\begin{equation}
\int_\cb\partial_i\left(\frac{\partial\ch}{\partial(\partial_i\phi)}\delta\phi\right) d^d r=-\int_\cs \frac{\partial\ch}{\partial(\partial_i\phi)}\delta\phi\,n_i\,d^{d-1} r\,.
\end{equation}

We may now calculate the force $\bff$, as defined in Eq.~(\ref{def_force}), in a given microstate of the fluid medium. We will examine successively the two different routes presented in the introduction, which correspond to two different ways of keeping $\phi$ constant.

\subsection{First route}

The first way we may proceed is to keep $\phi$ constant at each point in space during the infinitesimal transformation, or, in other words, to keep the field $\phi$ constant in the Eulerian sense. This amounts to taking $\delta\phi\equiv0$ in Eq.~(\ref{vartot}), which gives:
\begin{equation}
\bff ^{(1)}=-\be_i\left.\frac{\partial H}{\partial \ell_i}\right|_{\phi,\,\mathrm{Eul.}}=\int_\cs \left(\ch n_i-\frac{\partial V}{\partial \phi} \partial_i\phi \right) \,d^{d-1} r\,\,\be_i \,.
\label{prece}
\end{equation}
In the case where the coupling potential is $V(\phi,\bm{\nabla}\phi)$ instead of $V(\phi)$, a term $-(\partial V/\partial (\partial_j\phi)) \partial_i\partial_j \phi$ has to be added in the brackets in Eq.~(\ref{prece}).

However, it is physically not clear why each point in space should keep the same value of $\phi$ during a displacement in which each fluid particle of the system moves by $\bdr$. Another more formal argument also shows that it is artificial to keep the function $\phi$ constant while moving infinitesimally the inclusion: the domain of definition of $\phi$ itself depends on the position of the inclusion. For instance, when the inclusion is moved from $\ca$ (centered on $\bl$) to $\ca'$ (centered on $\bl+\bdl$), the initial $\phi$ is not defined in the region $\ca\smallsetminus\ca'$ where it should exist after the transformation. One way to deal with this mathematical issue is to consider the analytic continuation of $\phi$ in this small region, and to truncate $\phi$ in $\ca'\smallsetminus\ca$, but the physical meaning of this process is unclear. 

In other words, this first route is not adapted to calculate the force on an embedded inclusion because the position $\bl$ of the inclusion and the Eulerian field $\phi$ are not independent variables. Let us now move on to the second route.

\subsection{Second route}
Let us consider the example of critical binary mixtures, where $\phi$ is the (shifted) concentration: during a displacement that is smooth at the microscopic scale, each fluid particle keeps the same concentration, so $\phi$ is transported by fluid particles. Similarly, in the case of liquid crystals, the order parameter field is transported. The case of membranes and interfaces is a little bit more complex (see the discussion in Sec.~\ref{disc}). 

Let us focus on fields that are transported by fluid particles during a deformation. For such a field $\phi$, the correct force $\bff$ will be provided by the second route defined in the introduction, where $\phi$ is kept constant in the Lagrangian sense, i.e., each fluid particle of the medium keeps a constant $\phi$ during the infinitesimal transformation. In this case, $\delta^T\!\phi\equiv0$, so that $\delta\phi(\br)=-\bm{\nabla}\phi(\br)\cdot\bdr(\br)$ for all $\br$ in $\cb$: Eq.~(\ref{vartot}) becomes
\begin{align}
\delta H=&-\int_\cb\left[\frac{\partial\ch}{\partial\phi}-\partial_j\left(\frac{\partial\ch}{\partial(\partial_j\phi)}\right)\right]\partial_i\phi\,\delta r_i\,d^{d} r\nonumber\\
&-\delta\ell_i\int_\cs\left[\ch\delta_{ij}-\frac{\partial\ch}{\partial(\partial_j\phi)}\partial_i\phi\right]n_j\,d^{d-1} r\,,
\label{varlagr}
\end{align}
Using the stress tensor $\bT$ of the fluid medium, which is discussed and derived in Appendix~\ref{T_der}, and which reads
\begin{equation}
T_{ij}=\ch\delta_{ij}-\frac{\partial\ch}{\partial(\partial_j\phi)}\partial_i\phi\,,
\label{expr_T}
\end{equation}
and its divergence
\begin{equation}
\partial_j T_{ij}=\left[\frac{\partial\ch}{\partial\phi}-\partial_j\left(\frac{\partial\ch}{\partial(\partial_j\phi)}\right)\right]\partial_i\phi=\frac{\delta H}{\delta\phi}\partial_i\phi\,,
\label{divT}
\end{equation}
we can rewrite Eq.~(\ref{varlagr}) as
\begin{equation}
\delta H=-\int_\cb\partial_j T_{ij}\,\delta r_i\,d^{d} r-\delta\ell_i\int_\cs T_{ij} n_j\,d^{d-1} r\,.
\label{deltaH}
\end{equation}
This relation being valid for any continuous displacement field $\bdr$ such that $\forall \br\in\ca,\,\,\bdr=\bdl$, it yields the force $\bff^{(2)}$ exerted by the medium on the inclusion, and the hypervolume density $\bq$ of internal forces in the fluid medium. Indeed, we can carry out an identification with Eq.~(\ref{bilan_incl}), which comes from the principle of virtual work (see Appendix \ref{sec_def}). We obtain
\begin{align}
&\bff^{(2)}=-\be_i\left.\frac{\partial H}{\partial \ell_i}\right|_{\phi,\,\mathrm{Lagr.}}=\int_\cs T_{ij} n_j\,d^{d-1} r\,\be_i\,,\label{f2_g}\\
&\forall\br\in \cb,\,\,\bq(\br)=-\frac{\delta H}{\delta r_i (\br)}\be_i=\partial_j T_{ij}(\br)\,\be_i\,.
\label{q_tc}
\end{align}

This second method of calculating $\bff$ gives the integral of the stress tensor of the fluid medium on the boundary $\cs$ of the inclusion. Contrary to $\bff^{(1)}$, the force $\bff^{(2)}$ does not depend on the short-range coupling $V$ between the inclusion and the medium (in particular, Eq.~(\ref{f2_g}) holds for $V(\phi,\bm{\nabla}\phi)$ as well as for $V(\phi)$). This comes from the continuity of the function $\bdr$: the fluid particles of the medium that are infinitesimally close to the inclusion undergo the same small displacement as the inclusion, so the short-range interaction between the medium and the inclusion does not vary during the displacement. In a real displacement in a viscous fluid, the displacement field has to be continuous at the inclusion boundary, and we consider virtual displacements consistent with this constraint. Note that although the force $\bff^{(2)}$ in each microstate does not depend on $V$, the usual Casimir-like force, which is the thermal average of $\bff^{(2)}$ (see Sec.~\ref{Casi}), can depend on $V$ since the statistical weight of each microstate depends on $V$.

Thus, the right definition of the force exerted by the medium on an embedded inclusion, obtained via the second route, is given by the integral of the stress tensor of the medium. This corresponds to the definition used in Refs.~\cite{Bartolo02, Bartolo03, Najafi04, Gambassi06, Gambassi08, Bitbol10, Rodriguez11}, and it is also very close to the definition used in Refs.~\cite{Brito07,Buenzli08}, which is based on integrating a density-dependent pressure on the boundary of the inclusion. 

\subsection{Domain of application}
\label{disc}
We have considered a Hamiltonian density $\ch$ and a coupling potential $V$ depending on $\phi$ and $\bm{\nabla}\phi$. More generally, $\ch$ and $V$ can depend on higher-order derivatives of $\phi$. Our work can be adapted to such cases. In particular, the force $\bff^{(2)}$ exerted on an inclusion can be expressed as the integral of the stress tensor of the medium also in these cases. The stress tensor associated with a Hamiltonian (or Lagrangian) density involving higher-order derivatives is discussed, e.g., in Refs.~\cite{Podolsky44, Tielheim67}.

We have focused on fields that are transported by fluid particles during a displacement, such as the order parameter field in a critical mixture or in a liquid crystal. However, our reasoning can be adapted to other cases, especially when the field $\phi$ has a direct relation to the position of the fluid particle. For instance, the Hamiltonian density of a membrane described in the Monge gauge depends on the height $\phi(x,y)$ of the membrane with respect to a reference plane: for a displacement $\bm{\delta r} (x,y)$ of the fluid particle initially in $(x,y)$, the variation of the height $\phi$ at point $(x,y)$ is $\delta \phi=\delta r_z-\partial_x \phi\,\delta r_x-\partial_y \phi\,\delta r_y$ instead of $\delta\phi=-\partial_i\phi\,\delta r_i$. A stress tensor can be defined for the membrane, taking this particularity into account \cite{Capovilla02, Fournier07, Futur}. Note that lipid membrane models (e.g., the Helfrich model \cite{Helfrich73}) involve the membrane curvature, and thus, second derivatives of the height $\phi$. We showed in Ref.~\cite{Bitbol10} that the thermal average of the integral of the membrane stress tensor around a pointlike inclusion, in presence of a second inclusion, gives the usual Casimir-like force between the two inclusions, and we also used the membrane stress tensor to study the fluctuations of this force. 

In spite of this wider domain of application, the present work is restricted to the case of a fluid medium described by a scalar field $\phi$ such that its change during an infinitesimal displacement is a function of this infinitesimal displacement. More precisely, the relation $\delta\phi=-\partial_i\phi\,\delta r_i$ (or its equivalent for the membrane) is crucial in our derivation of $\bff^{(2)}$. As a counterexample, let us consider a one-dimensional elastic solid described by a scalar deformation field $\phi(X)=x(X)-X$, where $X$ is the position of a particle in the reference (nondeformed) configuration of the solid, while $x(X)$ is its position in the deformed configuration considered. In this case, changing the reference coordinate $X$ of a particle has no link with changing the deformation field $\phi$. 

In addition, the assumption that the displacement $\bdr$ is continuous at the boundary of the inclusion is valid for a viscous fluid, however small its viscosity, but not for an ideal fluid. In the latter case, only the component of the displacement normal to the inclusion boundary has to be continuous. Thus, our approach may not be fully adapted to a superfluid. However, it is adapted to study Casimir-like forces in a superfluid in the parallel plate geometry, which is the one usually considered. Indeed, modifying the distance between the plates involves a displacement perpendicular to the plates. 

\subsection{Mean-field configuration}
\label{mf1}

The quantities $\bff^{(1)}$ and $\bff^{(2)}$ obtained via the two different routes are, in general, not equal. However, they cannot be distinguished in the most probable configuration of the field $\phi$, i.e., in the mean-field configuration. Indeed, this configuration is such that the energy $H$ is stationary with respect to variations of $\phi$:
\begin{align}
0=\delta H&=\int_\cb\left[\frac{\partial\ch}{\partial\phi}-\partial_i\left(\frac{\partial\ch}{\partial(\partial_i\phi)}\right)\right]\delta\phi\,d^{d} r\nonumber\\
&+\int_\cs\left[\frac{\partial V}{\partial \phi}-\frac{\partial\ch}{\partial(\partial_i\phi)}n_i\right]\delta\phi\,d^{d-1} r\,,
\label{var_phi}
\end{align}
for any $\delta\phi$.
The bulk equilibrium condition yields the usual Euler-Lagrange equation, valid on $\cb$: 
\begin{equation}
\frac{\partial\ch}{\partial\phi}-\partial_i\left(\frac{\partial\ch}{\partial(\partial_i\phi)}\right)=0\,,
\end{equation}
while the boundary equilibrium condition gives the following relation, valid on $\cs$:
\begin{equation}
\frac{\partial V}{\partial \phi}=\frac{\partial\ch}{\partial(\partial_i\phi)}n_i\,.
\label{boundary}
\end{equation}
When Eq.~(\ref{boundary}) holds, $\bff^{(1)}$ and $\bff^{(2)}$ are identical. Thus, the difference between the two routes is irrelevant when one considers the mean-field configuration.

\section{Non-embedded influencing object}
\label{oio}

\begin{figure}[h t b]
\centerline{\includegraphics[width=8cm]{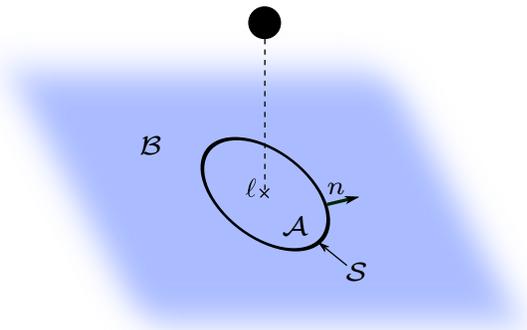}}
\caption{(Color online) Schematic representation of a two-dimensional fluid medium with an influencing object. The object, represented by a black sphere, is not embedded in the medium (here, it is above the plane where the medium stands). It influences the medium, i.e., the field $\phi$, in a region $\ca$, which is centered on $\bl$. The other notations are the same as in Fig.~\ref{emb}.}
\label{weak}
\end{figure}

Let us now study the case of an object that interacts with the fluid medium without being embedded in it (see Fig.~\ref{weak}). For instance, it may be an optical trap creating a local electromagnetic field in the medium, or a protein binding very softly onto a lipid membrane so that the latter keeps its fluidity: in these cases, there is no material object immersed in the medium. We will refer to this type of object as an ``influencing object'', to make the distinction with the embedded inclusion. We will see that, contrary to the case of the embedded inclusion, it is here the first route that gives the correct force $\bff$.

In the case of the influencing object, the field $\phi$ with Hamiltonian density $\ch$ is defined everywhere in $\mathbb{R}^d$ (recall that the $d$-dimensional fluid medium is assumed to be infinite). In the region $\ca$, the medium is affected by the influencing object: a term $V(\phi)$, representing the interaction between the influencing object and the medium, adds to $\ch$. The effective Hamiltonian $H$ of the fluid medium with the influencing object reads
\begin{align}
H&=\int_{\mathbb{R}^d}\ch(\phi,\bm{\nabla}\phi)\,d^d r
+\int_\ca V(\phi)\,d^d r\nonumber\\
&=\int_{\mathbb{R}^d}\left[\ch+V\,\bm{1}_\ca\right] \,d^d r\,,
\label{E_transp}
\end{align}
where $\bm{1}_\ca$ is the indicator function of the region $\ca$. Here too, our work can be generalized to $V(\phi,\bm{\nabla}\phi)$, and the final results for the force will be mentioned in this case.

In order to calculate the force $\bff$ defined in Eq.~(\ref{def_force}), we can follow the same path as in the case of the embedded inclusion, by applying the generic transformation (\ref{trf_inf}). The variation of $H$ during this transformation reads, at first order:
\begin{align}
\delta H&=\int_{\mathbb{R}^d}\frac{\delta H}{\delta\phi} \,\delta\phi\,d^d r+\int_\cs V(\phi)\,n_i\,\delta\ell_i\,d^{d-1} r\,,
\label{vartot_transp}
\end{align}
where the functional derivative of $H$ with respect to $\phi$ is
\begin{equation}
\frac{\delta H}{\delta\phi}=\frac{\partial\ch}{\partial\phi}+\frac{\partial V}{\partial\phi}\bm{1}_\ca-\partial_i\left(\frac{\partial\ch}{\partial(\partial_i\phi)}\right)\,.
\label{derfct}
\end{equation}

\subsection{First route}

If $\phi$ is kept constant in the Eulerian sense during the infinitesimal displacement (i.e., $\delta\phi\equiv0$), we obtain
\begin{equation}
\bff^{(1)}=-\be_i\left.\frac{\partial H}{\partial \ell_i}\right|_{\phi,\,\mathrm{Eul.}}=-\int_\cs V(\phi)\,\bm{n}\,d^{d-1} r\,.
\label{f1}
\end{equation}
The result for $V(\phi,\bm{\nabla}\phi)$ is exactly similar.

In fact, since the influencing object is not embedded in the medium, the force exerted on it by the medium in a given microstate can be calculated directly by moving the object \emph{with respect to the medium} in a given configuration of $\phi$. In other words, $\bl$ and the Eulerian field $\phi$ can be considered as independent variables. Then, we only need to take into account the variation of the interaction energy $E_p(\bl)=\int_\ca V(\phi)\,d^d r$ between the medium and the object when the position $\bl$ of the object is modified:
\begin{equation}
\bff^{(1)}=-\be_i\left.\frac{\partial E_p}{\partial \ell_i}\right|_{\phi,\,\mathrm{Eul.}}=-\int_\cs V(\phi)\,\bm{n}\,d^{d-1} r\,.
\end{equation}
This derivation is more physical than considering the full variation of $H$ during the generic transformation (\ref{trf_inf}). 
It shows that here, $\bff^{(1)}$ is simply the negative gradient of the potential energy of interaction $E_p$ between the medium and the influencing object. 

This definition of the force, which is the correct one for influencing objects that are not embedded in the medium, is the one used in Refs.~\cite{Dean09, Dean10, Demery10a, Demery10b}. In these works, this definition of the force is used to investigate Casimir-like interactions out of equilibrium \cite{Dean09, Dean10}. Note however that Casimir-like interactions are usually studied between embedded inclusions and not between non-embedded influencing objects. Besides, this definition has also been used to investigate drag forces in classical fields \cite{Demery10a, Demery10b}. In the latter works, the example studied in detail is a pointlike magnetic field moving through an Ising ferromagnet: it qualifies as an influencing object, since nothing material is embedded in the ferromagnet. However, the present work shows that the application of these results to the diffusion of inclusions embedded in membranes, which is discussed in Refs.~\cite{Demery10a, Demery10b}, should be taken with caution.

\subsection{Second route}

The Hamiltonian in Eq.~(\ref{E_transp}) can be used to describe a medium coupled to an influencing object, but also a medium containing a ``perturbative embedded inclusion'', i.e., an inclusion that is only slightly different from the surrounding medium. Indeed, in the latter case, $\phi$ is defined everywhere in $\mathbb{R}^d$, and the energy density is perturbed by an extra term $V$ inside the inclusion. For instance, in lipid membranes, domains with a lipid composition different from that of the rest of the membrane can be described as perturbative embedded inclusions, as well as similar structures in nematic liquid crystals. 

As for any other inclusion, it is meaningless to move a perturbative inclusion while keeping $\phi$ constant in the Eulerian sense, since moving the inclusion displaces the surrounding fluid particles. If we assume, as before, that $\phi$ is transported by fluid particles, the second route is the right one to calculate the force exerted on a perturbative embedded inclusion. Let us now calculate this force.

If the field $\phi$ is kept constant in the Lagrangian sense during the infinitesimal transformation, i.e., $\delta\phi(\br)=-\bm{\nabla}\phi(\br)\cdot\bdr(\br)$ for all $\br$ in $\mathbb{R}^d$, Eq.~(\ref{vartot_transp}) becomes
\begin{align}
\delta H=&-\int_\cb \frac{\delta H}{\delta\phi}\, \partial_i\phi\,\delta r_i\,d^{d} r\nonumber\\
&-\left\{\int_\ca \frac{\delta H}{\delta\phi}\,\partial_i\phi\,d^{d} r-\int_\cs V(\phi)\,n_i\,d^{d-1} r\right\}\delta\ell_i\,,
\label{varlagr_transp}
\end{align}
where, again, $\cb=\mathbb{R}^d\smallsetminus\ca$. This yields, using Eq.~(\ref{derfct}),
\begin{align}
\bff^{(2)}=&-\be_i\left.\frac{\partial H}{\partial \ell_i}\right|_{\phi,\,\mathrm{Lagr.}}\nonumber\\
=&\int_\ca \frac{\delta H}{\delta\phi}\bm{\nabla}\phi\,d^{d} r-\int_\cs V(\phi)\,\bm{n}\,d^{d-1} r\nonumber\\
=&\int_\ca\left[\frac{\partial\ch}{\partial\phi}+\frac{\partial V}{\partial\phi}-\partial_j\left(\frac{\partial\ch}{\partial(\partial_j\phi)}\right)\right]\bm{\nabla}\phi\,d^{d} r\nonumber\\
&-\int_\cs V(\phi)\,\bm{n}\,d^{d-1} r\,.
\label{f2_1}
\end{align}
Given the singularities in the energy density $\ch+V\,\bm{1}_\ca$ on $\cs$, the integral over $\ca$ has to be calculated using the procedure defined in Appendix~\ref{sec_def}, Eq.~(\ref{procedure}). 
Since $(\partial V/\partial\phi)\bm{\nabla}\phi=\bm{\nabla}V$, the two terms involving $V$ in Eq.~(\ref{f2_1}) cancel. Thus, using Eq.~(\ref{divT}), we obtain
\begin{equation}
\bff^{(2)}=\int_\ca\partial_j T_{ij}\,d^{d} r\,\be_i=\int_\cs T_{ij} n_j\,d^{d-1} r\,\be_i\,,
\end{equation}
where $\bm{T}$ is the stress tensor of the fluid medium.
This result, which is independent of $V$, remains the same for $V(\phi,\bm{\nabla}\phi)$.

The expression for $\bff^{(2)}$ is the same here as for the embedded inclusion. In fact, a perturbative inclusion is a particular inclusion with $E_\mathrm{incl}=\int_\ca (\ch+V)\,d^d r$, which verifies $\delta E_\mathrm{incl}=0$ during our transformation, and without any explicit boundary coupling. 

Thus, while it is the first route that gives the correct force exerted by the medium on an influencing object, the second route is the right one in the case of a perturbative embedded inclusion. 

\subsection{Mean-field configuration}

The two forces $\bff^{(1)}$ and $\bff^{(2)}$ are, in general, not equal, and we have seen that they are relevant to very different physical situations. However, as in the case of an embedded inclusion (see Sec.~\ref{mf1}), these two quantities are equal in the mean-field configuration of the system. Indeed, in this configuration, $\delta H/\delta\phi\equiv0$, so Eq.~(\ref{f2_1}) shows that $\bff^{(2)}$ is identical to $\bff^{(1)}$.  

\subsection{Formal relation between the two types of forces}

Independently of the physical interpretations of the forces $\bff^{(1)}$ and $\bff^{(2)}$, Eq.~(\ref{f2_1}) gives a formal relation between these two forces:
\begin{equation}
\bff^{(1)}=\bff^{(2)}-\int_\ca\frac{\delta H}{\delta\phi}\bm{\nabla}\phi\,d^{d} r\,.
\label{relf}
\end{equation}
Given the singularities in the energy density $\ch+V\,\bm{1}_\ca$ on $\cs$, the integral over $\ca$ has to be calculated using the procedure defined in Appendix~\ref{sec_def}, Eq.~(\ref{procedure}). 

Let us introduce the stress tensor $\bT'$ of the composite medium comprising the perturbative embedded inclusion: the force density at each point of the medium, even inside the perturbative embedded inclusion, is $\bq'=\partial_j T'_{ij} \be_i$ (see Appendix~\ref{Ttilde_der}). Thus, the basic definition Eq.~(\ref{for_dens}) of the force $\bff$ enables us to write:
\begin{equation}
\bff^{(2)}=\int_{\ca} \bq'(\br)\,d^d r\,.
\end{equation}
Here again, the integral over $\ca$ has to be calculated using the procedure defined in Eq.~(\ref{procedure}). 
Besides, if $\cc$ is a region such that $\ca \subset \cc$, we have
\begin{align}
\int_\cc \bq'(\br)\,d^d r&=\int_\cc\frac{\delta H}{\delta\phi}\bm{\nabla}\phi\,d^d r-\int_\cs V(\phi)\,\bm{n}\,d^{d-1} r\,,
\end{align}
where we have used the expression of $\bq'$ in Eq.~(\ref{qp}).
Thus, we can write
\begin{equation}
\bff^{(1)}=\int_\cc \bq'(\br)\,d^d r-\int_\cc\frac{\delta H}{\delta\phi}\bm{\nabla}\phi\,d^d r\,,
\end{equation}
which corresponds exactly to formula (18) in \cite{Dean10}.
However, this equation relates $\bff^{(1)}$ and $\bff^{(2)}$ only if $\cc = \ca$: the formal relation between $\bff^{(1)}$ and $\bff^{(2)}$ is given by Eq.~(\ref{relf}).

\section{Casimir-like force}
\label{Casi}

\subsection{Embedded inclusions}
Let us consider a fluid medium comprising two embedded inclusions with respective centers of mass at the origin of the frame and at point $\bl$, and let us introduce the unit vector $\bm{u}\equiv\bl/\ell$. 
The Casimir-like force exerted on the inclusion in $\bl$ by the other one is usually defined through $\bff_C=-\bm{u}\,\partial F/\partial \ell$, where $F(\ell)=-k_\mathrm{B}T\ln Z(\ell)$ is the free energy of the system. 
We are going to show that $\bff_C$ corresponds to the \emph{average at thermal equilibrium} of the force $\bff^{(2)}$ exerted on the inclusion in $\bl$ by the medium containing the other inclusion.

Let us assume that the line joining the centers of mass of the two inclusions is a symmetry axis of the system. This assumption is valid in the standard case of parallel plates, as well as for pointlike and spherical inclusions. Then, the effective Hamiltonian $H$ of the system only depends on $\phi$ and on the distance $\ell$ between the two inclusions.  
At thermal equilibrium, the statistical weight of a configuration $([\phi],\ell)$ is given by $e^{-\beta H([\phi],\ell)}$, where the notation $[\phi]$ represents a functional dependence on $\phi$. At a given $\ell$, the partition function of the system is
\begin{equation}
Z(\ell)=\int\!\mathcal{D}\phi\,\,e^{-\beta H([\phi],\ell)}\,,
\label{Zl}
\end{equation}
where the functional integral runs over the functions $\phi$ defined in the domain $\cb=\mathbb{R}^d\smallsetminus(\ca_{\bm{0}}\cup\ca_{\bl})$, where $\ca_{\bl}$ is the region where the inclusion with center of mass in $\bl$ stands (and similarly for $\ca_{\bm{0}}$).

Let us now introduce a replica of the system presented above, identical to it except that the center of mass of the second inclusion is at point $\bl+\delta\ell\,\bm{u}$. The partition function of this replica reads
\begin{equation}
Z(\ell+\delta\ell)=\int\!\mathcal{D}'\tilde\phi\,\,e^{-\beta H([\tilde\phi],\ell+\delta\ell)}\,.
\label{Zldl}
\end{equation}
Here, the functional integral runs over the functions $\tilde\phi$ defined in the domain $\cb'=\mathbb{R}^d\smallsetminus (\ca_{\bm{0}} \cup \ca_{\bl+\delta\ell\bm{u}})$. This difference with respect to Eq.~(\ref{Zl}) is symbolized by a prime on the functional measure. Let us consider a smooth invertible function $\br':\cb\to\cb',\,\br\mapsto\br'(\br)=\br+\bdr(\br)$, where the infinitesimal virtual displacement field $\bdr$ is such that $\bdr=\bm{0}$ on the first inclusion while $\bdr=\delta\ell\,\bm{u}$ on the second one, and $|\bdr|/\ell$ is small, say of order $\epsilon$. Let us assume that the function $\br'$ maps a field $\phi$ to a field $\tilde\phi=\phi+\delta\phi$ such that $\tilde\phi(\br'(\br))=\phi(\br)$ for all $\br\in\cb$, and $\delta\phi=-\bdr\cdot\bm{\nabla}\phi$ at first order in $\epsilon$. This is motivated by our assumption that each fluid particle of the medium keeps the same value of $\phi$ during any real smooth infinitesimal displacement. As this process can be inverted, $\br'$ yields a one-to-one mapping of the states $([\phi],\ell)$ of the first system to the states $([\phi+\delta\phi],\ell+\delta\ell)$ of the replica. We may thus write
\begin{equation}
\mathcal{D}\phi=\prod_{\br\in \cb} d[\phi(\br)]=\prod_{\br'(\br)\in \cb'} d[\tilde\phi(\br'(\br))]=\mathcal{D}'\tilde\phi\,,
\end{equation}
where the continuous products must be understood as $\prod_{\br\in \cb} d[\phi(\br)]\equiv \lim_{N\to\infty} \prod_{n=1}^N d [\phi(\br_n)]$, where $\{\br_n\}$ is a regular mesh of $\cb$ \cite{Schulman}. Hence, Eq.~(\ref{Zldl}) can be rewritten as
\begin{equation}
Z(\ell+\delta\ell)=\int\!\mathcal{D}\phi\,\,e^{-\beta H([\phi+\delta\phi],\ell+\delta\ell)}\,,
\end{equation}
so the difference of free energy between the replica and the original system reads at first order in $\epsilon$:
\begin{equation}
\delta F=-k_\mathrm{B} T\,\frac{\delta Z}{Z}=\int \mathcal{D}\phi\,\,\frac{e^{-\beta H([\phi],\ell)}}{Z}\,\delta H=\langle \delta H\rangle\,,
\end{equation}
where $\langle.\rangle$ denotes the average at thermal equilibrium, while $\delta Z\equiv Z(\ell+\delta\ell)-Z(\ell)$, and $\delta H\equiv H([\phi+\delta\phi],\ell+\delta\ell)-H([\phi],\ell)$. 

The expression of $\delta H$ is given by Eq.~(\ref{deltaH}), so we obtain
\begin{equation}
\delta F=-\int_\cb\langle q_i\rangle\,\delta r_i\,d^{d} r-\delta\ell \int_\cs \langle T_{ui}\rangle n_i\,d^{d-1} r\,,
\label{dF_int}
\end{equation}
where the axes have been chosen so that one of them is along $\bm{u}$: $T_{ui} n_i\,d^{d-1} r$ denotes the component along this axis of the force transmitted through the infinitesimal hypersurface $d^{d-1} r$. Recall that $\bq$ represents the hypervolume density of internal in the medium, given by Eq.~(\ref{q_tc}).

As no average external forces are imposed to the system, the hypervolume density of external forces $\bw$ verifies $\langle \bw \rangle=\bm{0}$. Neglecting inertia, which is possible if the timescales considered are sufficiently large, Newton's second law applied to each fluid particle of the medium gives $\bq=-\bw$ (see Appendix~\ref{sec_def}). Thus, we obtain $\langle \bq \rangle=\bm{0}$. This relation can also be obtained formally: Eqs.~(\ref{divT}) and (\ref{q_tc}) show that $\bm{q}=(\bm{\nabla}\phi)\,\delta H/\delta\phi $, and the thermal average of the latter quantity can be proved to vanish using the Schwinger-Dyson equation \cite{Dean10}. Thus, Eq.~(\ref{dF_int}) becomes
\begin{equation}
\delta F=-\delta\ell \int_\cs \langle T_{ui}\rangle n_i\,d^{d-1} r=-\delta\ell\,\langle\bff^{(2)}\rangle\cdot\bm{u}\,,
\label{last_df}
\end{equation}
where we have used the expression of $\bff^{(2)}$ in Eq.~(\ref{f2_g}). Note that Eq.~(\ref{last_df}) is independent of the virtual displacement field $\bdr$ chosen at the beginning of this discussion to map the states of the system onto the state of its replica. We obtain from Eq.~(\ref{last_df}):
\begin{equation}
\bff_C=-\frac{\partial F}{\partial\ell}\,\bm{u}=\left(\langle\bff^{(2)}\rangle\cdot\bm{u}\right)\bm{u}\,.
\end{equation}
Since $\langle\bff^{(2)}\rangle$ is along $\bm{u}$ by symmetry, we can conclude that $\bff_C=\langle\bff^{(2)}\rangle$.

\subsection{Non-embedded influencing objects}
Although Casimir-like forces are usually studied between embedded inclusions, let us now consider non-embedded influencing objects. In this case, the partition function can also be written as
\begin{equation}
Z(\ell)=\int\!\mathcal{D}\phi\,\,e^{-\beta H([\phi],\ell)}\,,
\end{equation}
but here the functional integral runs over the functions $\phi$ defined on $\mathbb{R}^d$. In contrast to the case of the embedded inclusions, $\ell$ can be varied at constant Eulerian field $\phi$. Thus we can write directly
\begin{equation}
\frac{\partial F}{\partial \ell}=\int \mathcal{D}\phi\,\,\frac{e^{-\beta H([\phi],\ell)}}{Z}\,\frac{\partial H}{\partial\ell}=\left\langle\frac{\partial H}{\partial\ell}\right\rangle=-\langle\bff^{(1)}\rangle\cdot\bm{u}\,.
\end{equation}
Thus, in the case of influencing objects, it is the average of $\bff^{(1)}$ that gives $-\bm{u}\,\partial F/\partial\ell$. 

Some effective Hamiltonians, such as the one in Eq.~(\ref{E_transp}) and the one studied in the next section, can describe both influencing objects and embedded inclusions. For such effective Hamiltonians, our results show that the thermal average of $\bff$ is the same in these two physical cases, each of them being treated with the appropriate route. If there are two inclusions, this thermal average force corresponds to the Casimir-like force $-\bm{u}\,\partial F/\partial\ell$. Thus, the usual (i.e., thermal average) Casimir-like force does not depend on the route that is chosen. In spite of this degeneracy concerning thermal average, it is very important to distinguish the two physical cases as soon as one wishes to go beyond the thermal average force. This point will be illustrated in the next section.

\section{A simple example of Casimir-like force}

\subsection{The situation}

Let us now present a simple example with a Hamiltonian that can describe both influencing objects and embedded inclusions.
In this example, we calculate the variance of the Casimir-like force, both in the usual case of embedded inclusions (via the second route) and in the particular case of non-embedded influencing objects (via the first route), and we find two very different results. Thus, this example illustrates the importance of the distinction between the two physical cases.

Let us consider an infinite one-dimensional fluid medium described by a dimensionless scalar field $\phi$ with Hamiltonian density 
\begin{equation}
\ch=\frac{\kappa}{2} \phi'^2+\frac{m}{2}\phi^2\,,
\end{equation}
where $\phi'\equiv d\phi/dx$. The length scale $L=\sqrt{\kappa/m}$ which appears in $\ch$ corresponds to the correlation length of $\phi$, as shown in the following.
The energy $H_0$ of the medium is such that
\begin{align}
\beta H_0=\beta\int_\mathbb{R}dx\,\,\ch(x)=\frac{1}{2}\int_{\mathbb{R}^2}\!dx\,dy\,\,\phi(x)\co(x,y)\phi(y)\,,
\end{align}
where $\beta=(k_\mathrm{B}T)^{-1}$, and the symmetric operator $\co$ is defined by
\begin{align}
\co(x,y)=\left[\beta m-\beta\kappa\frac{d^2}{dx^2}\right]\delta(x-y)\,.
\end{align}
Let us assume that there are two pointlike inclusions or non-embedded influencing objects, in $x=0$ and $x=\ell$, where $\ell>0$, and that their coupling to the field $\phi$ is given by
\begin{align}
\beta V=\frac{\alpha}{2}\left[\phi^2(0)+\phi^2(\ell)\right]\,.
\end{align}
Since $\phi$ is dimensionless, $\alpha$ is dimensionless too. When $\alpha\to\infty$, this quadratic coupling yields Dirichlet boundary conditions on the inclusions. Indeed, in this limit, the statistical weight $e^{-\beta (H_0+V)}$ of any configuration such that $\phi\neq 0$ on an inclusion goes to zero.

For the calculations that follow, let us assume that there is an external field $u$ conjugate to $\phi$:
\begin{align}
\beta H_\mathrm{ext}=-\int_\mathbb{R}dx\,u(x)\phi(x)\,.
\end{align}
In order to calculate the partition function of the system
\begin{equation}
Z[u]=\int \! \mathcal{D}\phi\,\, e^{-\beta (H_0+V+H_\mathrm{ext})}\,,
\end{equation}
let us carry out a Hubbard-Stratonovich transformation \cite{ChaikinLubensky}. Using the relation
\begin{equation}
 e^{-\beta V}=\frac{1}{2\pi\alpha}\int_{\mathbb{R}^2} d^2 v\, \exp\left[-\frac{v^2}{2\alpha}+i\left(v_1\,\phi(0)+v_2\,\phi(\ell)\right)\right]\,,
\end{equation}
where $(v_1,v_2)$ is a two-dimensional vector, and performing the Gaussian integration on $\phi$, we obtain
\begin{align}
Z[u]=\frac{Z_0}{2\pi\alpha}\int_{\mathbb{R}^2} &d^2 v\, \exp\bigg[-\frac{v^2}{2\alpha}\nonumber\\
&+\frac{1}{2}\int_{\mathbb{R}^2}dx\,dy\,\,S(x)\,G(x,y)\,S(y)\bigg]\,.
\label{Zinterm}
\end{align}
In this expression, $Z_0$ is the partition function of the medium with no inclusion or other influencing object, while $S(x)=u(x)+i \,v_1\,\delta(x)+i\,v_2\,\delta(x-\ell)$, and $G$ is the Green function of $\co$. The latter can be obtained using a Fourier transform since the medium is infinite and translation-invariant:
\begin{align}
G(x,y)&=\frac{k_\mathrm{B}T}{2\pi}\int_\mathbb{R} dq\,\frac{e^{i\,q(x-y)}}{\kappa\, q^2+m}\nonumber\\
&=\frac{k_\mathrm{B}T\,L}{2\,\kappa}\,\,\exp\left(-\frac{|x-y|}{L}\right)\,,
\label{Green}
\end{align}
where $L=\sqrt{\kappa/m}$.
In the absence of inclusions or other influencing objects, the correlation function of $\phi$ is $G$, so $L$ represents the correlation length of $\phi$. Therefore, we expect Casimir-like forces to be most important when the distance $\ell$ between the objects is such that $\ell\ll L$. 

Performing the Gaussian integration on $(v_1,v_2)$ in Eq.~(\ref{Zinterm}) yields
\begin{equation}
Z[u]=\frac{Z_0}{\alpha\sqrt{\det M}}\, \exp\left(\frac{1}{2}\int_{\mathbb{R}^2}dx\,dy\,\,u(x)\,C(x,y)\,u(y)\right)\,,
\label{Z}
\end{equation}
where $C(x,y)=G(x,y)-A^T(x)\,M^{-1}A(y)$, with $A^T(x)=\left( G(x,0)\,,\,\,G(x,\ell)\right)$ and
\begin{equation}
M=\left( \begin{array}{cc}
G(0,0)+\alpha^{-1}&G(0,\ell)\\
G(0,\ell)&G(\ell,\ell)+\alpha^{-1}
\end{array} \right)\,.
\end{equation}
It is straightforward to deduce the moments of the Gaussian variable $\phi(x)$ from Eq.~(\ref{Z}):
\begin{align}
\langle\phi(x)\rangle&=-\left(\frac{1}{Z}\frac{\delta Z}{\delta u(x)}\right)[u=0]=0\,,\\
\langle\phi(x)\phi(y)\rangle&=-\left(\frac{1}{Z}\frac{\delta^2 Z}{\delta u(x)\,\delta u(y)}\right)[u=0]=C(x,y)\,.
\label{phiphi}
\end{align}
Thus, the correlation function of the field $\phi$ in the presence of the two objects is 
\begin{align}
\langle\phi(x)\phi(y)\rangle-\langle\phi(x)\rangle\langle\phi(y)\rangle&=C(x,y)\nonumber\\
&=G(x,y)-A^T(x)\,M^{-1}A(y)\,.  
\label{corr}                                                     
\end{align}

\subsection{Average Casimir-like force}

The average Casimir-like force $\left\langle f\right\rangle$ between the two inclusions can be calculated by differentiating the free energy $F=-k_\mathrm{B}T\,\ln(Z[u=0])$ of the system, which can be obtained from Eq.~(\ref{Z}):
\begin{align}
\left\langle f\right\rangle&=-\frac{\partial F}{\partial \ell}=-\frac{k_\mathrm{B}T}{2}\,\frac{\partial (\ln\det M)}{\partial \ell}\nonumber\\
&=\frac{-(k_\mathrm{B}T)^3 \alpha ^2 L\,e^{-\frac{2\ell}{L}}}{(k_\mathrm{B}T)^2 \alpha ^2 L^2 \left(1-e^{-\frac{2\ell}{L}}\right) +4 \,k_\mathrm{B}T\, \alpha \, \kappa \,L  +4 \,\kappa ^2}\,.
\label{fmoy_gen}
\end{align}

Taking the hard-constraint limit $\alpha\to\infty$, which will be denoted in the following by a subscript index $h$, we obtain:
\begin{equation}
\left\langle f_h\right\rangle=\frac{k_\mathrm{B}T}{L\left(1-e^{\frac{2\ell}{L}}\right)}  \underset{\ell\ll L}{\sim} -\frac{k_\mathrm{B}T}{2\,\ell}\,.
\label{hard}
\end{equation}
As expected, the Casimir-like force vanishes when $\ell\gg L$: when the distance between the inclusions is much larger than the correlation length of $\phi$, one inclusion cannot feel the effect of the other one. We have emphasized the ``critical regime'' $\ell\ll L$, where the Casimir-like force is most important: it has a simple $\ell^{-1}$ power-law dependence.

Meanwhile, in the perturbative limit $\alpha\to 0$, which will be denoted in the following by a subscript index $p$, we obtain, to lowest order in $\alpha$:
\begin{equation}
\left\langle f_p\right\rangle=-\frac{(k_\mathrm{B}T)^3 \alpha ^2 L\,e^{-\frac{2\ell}{L}}}{4\, \kappa ^2} \underset{\ell\ll L}{\sim} -\frac{(k_\mathrm{B}T)^3 \alpha ^2\,\left(L-2\ell\right) }{4\, \kappa ^2} \,.
\label{soft}
\end{equation}
Here too, we have emphasized the critical regime $\ell\ll L$: this time, the leading term is independent of $\ell$, so the force is nearly constant in this regime.

\subsection{Variance of the Casimir-like force}

We are now going to illustrate the difference between embedded inclusions and non-embedded influencing objects by studying the variance of the Casimir-like force. The two routes to calculate the force $\bff$ are detailed in the particular case of pointlike embedded inclusions or non-embedded influencing objects in Appendix~\ref{pointlike}.

\subsubsection{Preliminary remark}

The variance we are going to study is that of the force $\bff$ defined at the coarse-graining level where the system is described by $([\phi],\bl)$. This force is averaged over the fundamental microstates ``$\mu\to([\phi],\bl)$'' that yield the coarse-grained field $\phi$ and the inclusion position $\bl$. Indeed, it is shown in Appendix~\ref{sec_def} that the force $\bff_\mu$ exerted by the medium on the inclusion in the fundamental microstate $\mu$ verifies
\begin{equation}
\langle\bff_\mu\rangle_{\mu\to([\phi],\bl)}=\bff([\phi],\bl)\,,
\end{equation}
where $\langle.\rangle_{\mu\to([\phi],\bl)}$ represents the statistical average at equilibrium over the fundamental microstates $\mu\to([\phi],\bl)$. Therefore, the variance $\Delta^2\bff$ of the force $\bff$ is smaller than $\Delta^2\bff_\mu$. 

The difference between $\Delta^2\bff_\mu$ and $\Delta^2\bff$ is due to fluctuation modes with wavelengths smaller than the cutoff $a$ of the field $\phi$: such modes are averaged in the coarse-graining procedure leading to $\bff$. Thus, we expect $\Delta^2\bff$ to decrease if $a$ increases. The small-wavelength fluctuation modes are generically the fastest to equilibrate, so above a certain time resolution, it is right to consider $\Delta^2\bff$ instead of $\Delta^2\bff_\mu$, and for even longer time resolutions, it is appropriate to consider larger values of $a$. This subtelty, which is linked to the coarse-grained nature of the force $\bff$, arises in all studies of fluctuations of Casimir-like forces \cite{Bartolo02, Bitbol10}.

\subsubsection{Embedded inclusions}
In the case of pointlike embedded inclusions, the force exerted on the inclusion in $\ell$ by the medium containing the other inclusion is obtained via the second route. It is given by Eq.~(\ref{f2_pct}), adapted to $d=1$: $f^{(2)}= T(\ell^+)-T(\ell^-)$.

Let us first show explicitly in the present example that the thermal average of $f^{(2)}$ gives back the Casimir force $\langle f\rangle$. Here, the stress tensor of the medium is 
\begin{equation}
T=\ch-\phi'\frac{\partial\ch}{\partial\phi'}=-\frac{\kappa}{2}\phi'^2+\frac{m}{2}\phi^2\,,
\label{T_ex}
\end{equation}
and Eq.~(\ref{phiphi}) enables to express its average as
\begin{align}
\left\langle T(x)\right\rangle&=-\frac{\kappa}{2}\,C_{xy}(x,x)+\frac{m}{2}\,C(x,x)\,,
\end{align}
where we have introduced the notation $C_{xy}(x,y)\equiv[\partial_x \partial_y C](x,y)$. Such notations will be used in the following.
This average can be calculated thanks to Eq.~(\ref{Green}) and (\ref{corr}). Note that $G_{xy}(x,x)$ has to be regularized using the short-distance cutoff $a$ of the theory: 
\begin{align}
G_{xy} (x,x)&\equiv\frac{k_\mathrm{B}T}{2\pi}\int_{-1/a}^{1/a} dq\,\frac{q^2}{\kappa\, q^2+m}\nonumber\\
&=\frac{k_\mathrm{B}T}{\pi\,\kappa }\left(\frac{1}{a}-\frac{ \text{Arctan}\left(\frac{L}{a}\right)}{L}\right)\,.
\label{ddGreen}
\end{align}
We then recover the result obtained in Eq.~(\ref{fmoy_gen}): $\left\langle f^{(2)}\right\rangle=\left\langle T(\ell^+)\right\rangle-\left\langle T(\ell^-)\right\rangle=\left\langle f\right\rangle$. Note that the cutoff-dependent term Eq.~(\ref{ddGreen}) vanishes when one calculates the difference between the average stress on each side of the inclusion. This is reminiscent of the calculation of the (average) quantum Casimir force, either from the energy \cite{Casimir48} or from the radiation pressure \cite{Milonni88}, where two infinite quantities subtract to give a finite force. However, in the variance, this cutoff-dependent term will no longer vanish. 

We may now proceed to calculate the variance $\Delta^2 f^{(2)}$ of $f^{(2)}$:
\begin{equation}
\Delta^2 f^{(2)}=K(\ell^-,\ell^-)+K(\ell^+,\ell^+)-2\,K(\ell^-,\ell^+)\,,
\end{equation}
where $K(x,y)\equiv\left\langle T(x)T(y)\right\rangle-\left\langle T(x)\right\rangle\left\langle T(y)\right\rangle$. Given the expression of the stress tensor in Eq.~(\ref{T_ex}), its correlation function $K$ involves quartic terms in $\phi$ (or $\phi'$). Since $\phi(x)$ is a centered Gaussian variable, we may use Wick's theorem to calculate $K$. It yields:
\begin{align}
\Delta^2 f^{(2)}=&\frac{\kappa^2}{2}\left[C^2_{xy}(\ell^-,\ell^-)+C^2_{xy}(\ell^+,\ell^+)-2\,C^2_{xy}(\ell^-,\ell^+)\right]\nonumber\\
&-\kappa\,m\big[C^2_x(\ell^-,\ell^-)+C^2_x(\ell^+,\ell^+)\nonumber\\
&\hspace{1cm}-C^2_x(\ell^-,\ell^+)-C^2_y (\ell^-,\ell^+)\big]\,.
\end{align}
This variance can be calculated from Eqs.~(\ref{Green}), (\ref{corr}) and (\ref{ddGreen}). Note that $G_x (x,x)$ and $G_y (x,x)$ can be obtained from a regularized Fourier transform:  
\begin{align}
G_x (x,x)=- G_y (x,x)\equiv\frac{k_\mathrm{B}T}{2\pi}\int_{-1/a}^{1/a} dq\,\frac{iq}{\kappa\, q^2+m}=0\,.
\label{dGreen}
\end{align}
The full analytical expression of $\Delta^2 f^{(2)}$ is quite heavy, so we will only present its limiting behaviors for large and small $\alpha$, and we will focus on the regime where $a<\ell\ll L$, since the Casimir effect is strongest in the critical regime $\ell\ll L$. More precisely, in each limit, we carry out expansions in the small parameter $a/L$ after setting $\ell=C a$ where $C$ is a constant. 

Taking the hard-constraint limit $\alpha\to\infty$, and then keeping only the leading order in $a/L$, we obtain:
\begin{equation}
\Delta^2 f_h^{(2)}=\Delta^2 f^{(2)}{}^\dagger+\Delta^2 f_h^{(2)}{}^\ddagger\,,
\label{varhard0}
\end{equation}
where the first term is independent of $\ell$ and $\alpha$:
\begin{equation}
\Delta^2 f^{(2)}{}^\dagger\simeq \frac{(k_\mathrm{B}T)^2}{\pi^2 a^2}\,,
\label{varhard1}
\end{equation}
while the second one arises from the inclusions:
\begin{equation}
\Delta^2 f_h^{(2)}{}^\ddagger\simeq -\frac{(k_\mathrm{B}T)^2}{\pi \,a\,\ell}\,.
\label{varhard2}
\end{equation}
If we consider inclusions imposing Dirichlet boundary conditions (i.e., $\phi(0)=\phi(\ell)=0$) instead of inclusions imposing a potential $V$, we recover the result in Eqs.~(\ref{varhard0}, \ref{varhard1}, \ref{varhard2}) for the force variance. In this case too, the average force can be obtained either directly from $F$ or by using the stress tensor, and it gives back $\langle f_h\rangle$ in Eq.~(\ref{hard}). 

In the perturbative limit $\alpha\to 0$, let us keep the two lowest orders in $\alpha$: the leading term corresponds to the situation where there is no inclusion, $\alpha=0$, so the effect of the inclusions appears in the subleading term. We obtain
\begin{equation}
\Delta^2 f_p^{(2)}=\Delta^2 f^{(2)}{}^\dagger+\Delta^2 f_p^{(2)}{}^\ddagger\,,
\label{varrr0}
\end{equation}
where the first term is the one in Eq.~(\ref{varhard1}), still at leading order in $a/L$, while
\begin{equation}
\Delta^2 f_p^{(2)}{}^\ddagger\simeq -\frac{(k_\mathrm{B}T)^3 \alpha}{\pi\,\kappa\, a }\left(1-\frac{\ell}{L}\right)\,,
\label{varrr}
\end{equation}
where we have kept the two leading orders in $a/L$ to see the $\ell$-dependence.
Note that these results can be obtained directly from a perturbative expansion of the stress tensor correlation function $K$ at first order in $\beta V$. Besides, using this perturbative method, it is straightforward to study the more general problem of inclusions with a finite size $s$: we have checked that it gives back the average force Eq.~(\ref{soft}) and the variance Eqs.~(\ref{varrr0}, \ref{varrr}) in the small-size limit.

Both in the hard-constraint limit and in the perturbative limit (and also in the intermediary regime), the variance of the force features a term $\Delta^2 f^{(2)}{}^\dagger$, expressed in Eq.~(\ref{varhard1}), which is independent of $\ell$ and $\alpha$. It corresponds to the variance of the zero-average fluctuating force exerted on a point of the medium, or on a single inclusion, by the rest of the medium, in the absence of any other inclusion. This term depends on the cutoff $a$, and it decreases when $a$ increases, in agreement with the preliminary remark above. The other terms depend on $\ell$, and this dependence has the same origin as the Casimir-like interaction itself: it comes from the constraints imposed by the inclusions on the fluctuations of $\phi$. They also depend on the intensity of the coupling constant $\alpha$, because they come from the constraints imposed by the inclusions. The presence of a term independent of $\ell$ and of smaller ones that depend on $\ell$ is similar to the results of Refs.~\cite{Bartolo02,Bitbol10}. 

\subsubsection{Other influencing objects}

In the case of pointlike influencing objects that are not embedded in the fluid medium, the force exerted on the object in $\ell$ by the medium is obtained via the first route. It is given by Eq.~(\ref{f1_pct}), for $d=1$:
\begin{align}
f^{(1)}=-\phi'(\ell)\frac{\partial V}{\partial\phi}(\phi(\ell))=-\alpha\,k_\mathrm{B}T\,\phi'(\ell)\,\phi(\ell)\,.
\end{align}

Eq.~(\ref{phiphi}) enables to express its average as
\begin{align}
\left\langle f^{(1)}\right\rangle=-\alpha\,k_\mathrm{B}T\,C_x(\ell,\ell)\,,
\end{align}
which can be calculated from Eq.~(\ref{Green}), (\ref{corr}) and (\ref{dGreen}), yielding once more the result obtained in Eq.~(\ref{fmoy_gen}): $\left\langle f^{(1)}\right\rangle=\left\langle f\right\rangle$. We have thus verified on the present example that all the definitions (i.e., the direct derivation from the free energy, and the two routes defined in the introduction) give the same result for the thermal average of the force.

In order to calculate the variance of $f^{(1)}$, we use Wick's theorem as above. It gives
\begin{align}
\Delta^2 f^{(1)}&=\alpha^2 (k_\mathrm{B}T)^2\left[C(\ell,\ell)\,C_{xy}(\ell,\ell)+2\,C^2_x(\ell,\ell)\right]\,.
\end{align}
This variance can be calculated from Eqs.~(\ref{Green}), (\ref{corr}), (\ref{ddGreen}) and (\ref{dGreen}). Here again, we will only present the limiting behaviors of $\Delta^2 f^{(1)}$, in the same regimes as for the embedded inclusion.

In the hard-constraint limit $\alpha\to\infty$, keeping the leading order in $\alpha$, and then keeping the leading order in $a/L$, we obtain:
\begin{equation}
\Delta^2 f_h^{(1)}\simeq\frac{(k_\mathrm{B}T)^3 \alpha }{\kappa \,a}\left(\frac{1}{\pi}-\frac{a}{4\,\ell}\right)\,.
\end{equation}

In the perturbative limit $\alpha\to 0$, if we keep the two lowest orders in $\alpha$ (the lowest order term does not depend on $\ell$, so we also keep the next one), we obtain:
\begin{equation}
\Delta^2 f_p^{(1)}\simeq\frac{(k_\mathrm{B}T)^4 \alpha ^2}{2\, \pi\,\kappa^2}\,\frac{L}{a}+\frac{(k_\mathrm{B}T)^5  \alpha ^3 L}{2 \,\pi\,\kappa ^3}\left(-\frac{L}{a}+\frac{\ell}{a}+\frac{\pi }{4}\right)\,,
\label{Deltaf1p}
\end{equation}
where we have kept only the leading orders in $a/L$, and also the subleading one in the $\alpha ^3$ term to see the $\ell$-dependence.
This result can also be found directly from a perturbative expansion of $\left\langle(f^{(1)})^2\right\rangle$ at second order in $\beta V$.
This variance diverges when $L\to\infty$, but the correlation function itself diverges in this limit (see Eq.~(\ref{Green})).

The present results are very different from the ones concerning the embedded inclusions. First, there is no term independent of the coupling constant $\alpha$ here, since, as explained above, the first route amounts to taking the derivative of the potential energy of interaction between the medium and the influencing object. Besides, here, the variance diverges as $\alpha$ in the hard-constraint limit $\alpha\to\infty$, while it converged in the case of the embedded inclusion. Here follows a physical explanation of this divergence: for each microstate $\phi$, let $r$ and $s$ be such that $|\phi(\ell)|=\alpha^r$ and $|\phi(0)|=\alpha^s$. For $\alpha\to\infty$, only the states such that $r\leq-1/2$ and $s\leq-1/2$ can keep a significant statistical weight: they are such that $\exp(-\beta V)\geq1/e$, while for the others, $\exp(-\beta V)\to 0$ exponentially when $\alpha\to\infty$. 
Let us consider the statistically significant states with highest $r$, i.e., the ones such that $r=-1/2$: they will yield the largest forces $f^{(1)}$. For these states, $(f^{(1)})^2\sim \alpha\,(k_\mathrm{B}T)^2 \phi'^2(\ell)$. Since there is no particular constraint on $\phi'(\ell)$, $(f^{(1)})^2$ typically scales as $\alpha$ in these states, which explains the behavior of the variance. Much more qualitatively, if $\alpha$ is infinite, the only allowed states are such that $\phi(\ell)=0$, but $\phi'(\ell)$ is generically nonzero, so $\phi(\ell+\delta\ell)\neq 0$: if it moved to this position, the object would have an infinite energy. Thus, it feels an infinite restoring force. While such contributions can cancel in the average, they add in the variance, which thus diverges with $\alpha$.

In this simple example, we have found very different results for the variance of the Casimir-like force for embedded inclusions and for non-embedded influencing objects. It illustrates the importance of distinguishing these two physical cases and the associated routes for calculating the force.

The out-of-equilibrium behavior of Casimir-like forces also depends strongly on the route that is taken. The approach to equilibrium after a quench at $\phi=0$ has been studied for plates imposing boundary conditions to the field $\phi$, using the second route in Ref.~\cite{Gambassi08}, and using the first route in Ref.~\cite{Dean10}. The two routes give very different results for the case where one plate imposes Dirichlet boundary conditions while the other one imposes Neumann boundary conditions. Indeed, in Ref.~\cite{Gambassi08}, the force exerted on one plate is not equal to the one exerted on the other plate during the relaxation (but both converge to the same equilibrium value), while in Ref.~\cite{Dean10}, these two forces are always equal. Note that obtaining different forces on each plate is not ruled out by the action-reaction principle, since the forces at stake are forces exerted by the medium containing one plate on the other plate, and not forces exerted by one plate on the other plate.

\section{Conclusion}
We have investigated the force exerted on an inclusion by a fluid medium with long-range correlations, described by a scalar field $\phi$ in a coarse-grained theory. If a second inclusion is embedded in the medium, the thermal average of this force gives the Casimir-like force between the two inclusions. In order to go beyond the thermal average force, it is necessary to define properly the force $\bff$ exerted on an inclusion by the medium in each microstate. In practice, one must take the negative gradient of the effective Hamiltonian with respect to the position of the inclusion in a given microstate of the medium. We have shed light onto the subtlety of this task, showing two routes that can be taken to calculate this gradient. In the first route, $\phi$ is kept constant in the Eulerian sense, while in the second one, $\phi$ is kept constant in the Lagrangian sense.

In the usual case of an embedded inclusion, the position of the inclusion and the Eulerian field $\phi$ are not independent variables, so one should not take the first route. Indeed, when an inclusion is displaced infinitesimally, the surrounding fluid particles are displaced too. In many physical cases, $\phi$ is transported by the fluid particles of the medium during a displacement, so the second route is the correct one. It gives the integral of the stress tensor of the medium on the boundary of the inclusion.

We have also considered the case of influencing objects that interact with the medium without being embedded in it. Contrary to inclusions, such objects can be moved with respect to the medium at a given Eulerian field $\phi$. Thus, the first route is adapted to calculate the force exerted by the medium on an influencing object. 

In a nutshell, the two formal routes for calculating the force $\bff$ apply to two different physical cases. We have discussed the practical importance of this distinction. First, in the mean-field configuration, the two routes give the same result. Then, for effective Hamiltonians that can describe both physical cases, the two routes give the same thermal average of the force. However, the difference between these two routes becomes crucial as soon as one wants to study this force beyond its mean-field value and its thermal average. We have shown in a simple example that they yield very different results for the variance of the Casimir-like force. 
Besides, comparing Refs.~\cite{Gambassi08} and~\cite{Dean10} shows that the out-of-equilibrium behavior of Casimir-like forces also depends on the route that is taken. 

Two definitions of the force, corresponding to our two routes, are currently used to study the fluctuations of the Casimir-like force and its out-of-equilibrium behavior. Our work shows that the second route, which gives the integral of the stress tensor, should be used when studying Casimir-like forces between embedded inclusions, in agreement with Refs.~\cite{Bartolo02, Bartolo03, Najafi04, Gambassi06, Gambassi08, Bitbol10, Rodriguez11}. In contrast, the first route, which is used in Refs.~\cite{Dean09, Dean10, Demery10a, Demery10b}, should be reserved to the case of non-embedded influencing objects.

\appendix

\section{Definition of the force $\bff$ from the principle of virtual work}
\label{sec_def}
The aim of the present Appendix is to provide a justification of the fundamental definition Eq.~(\ref{def_force}) in our coarse-grained, effective description. Let us consider, as in the main text, an infinite $d$-dimensional fluid medium with short-range interactions, and let us assume that an embedded inclusion extends over the region $\ca\subset\mathbb{R}^d$ of this medium. 
Let us consider a ``fundamental'' microstate $\mu$ of the system constituted by the fluid medium with the inclusion, i.e., a microstate defined not by the coarse-grained field $\phi$ and by the position $\bl$ of the center of mass of the inclusion, but by all the underlying microscopic degrees of freedom. Let us call $\bw_\mu$ the hypervolume density of forces exerted by the exterior on the system, and $\bq_\mu$ the hypervolume density of forces exerted on a piece of the system by the rest of the system, in the microstate $\mu$. Both $\bw_\mu$ and $\bq_\mu$ can take different values in the different fundamental microstates ``$\mu\to([\phi],\bl)$'' that yield the same coarse-grained field $\phi$ and inclusion position $\bl$. 
The coarse-graining procedure regroups these microstates so that
\begin{equation}
Z([\phi],\bl)=e^{-\beta H([\phi],\bl)}=\sum_{\mu\to([\phi],\bl)}e^{-\beta E_\mu}\,,
\end{equation}
where $E_\mu$ is the energy of the fundamental microstate $\mu$, while $H([\phi],\bl)$ is the effective energy of the coarse-grained microstate $([\phi],\bl)$.
Let us now introduce the average $\bq([\phi],\bl)$ (respectively, $\bw([\phi],\bl)$) of $\bq_\mu$ (respectively, $\bw_\mu$) over the microstates $\mu\to([\phi],\bl)$: $\bq$ and $\bw$ are coarse-grained force densities. Explicitly, we have
\begin{equation}
\bq([\phi],\bl)\equiv \langle\bq_\mu\rangle_{\mu\to([\phi],\bl)}=\sum_{\mu\to([\phi],\bl)}\bq_\mu \frac{e^{-\beta E_\mu}}{Z([\phi],\bl)}\,,
\label{cg_q}
\end{equation}
and similarly for $\bw$. 

Let us consider a material particle of hypervolume $d^d r$ of the system: it can be a fluid particle of the medium in the sense defined in the main text, or a piece of  the inclusion. Since it is a (macroscopically) closed particle, Newton's second law applies to it. In the fundamental microstate $\mu$, it reads $\bw_\mu\,d^d r+\bq_\mu\,d^d r=d(\bm{p}_\mu\,d^d r)/dt$, where $\bm{p}_\mu$ is the hypervolume density of momentum in the microstate $\mu$. This is equivalent to writing that $\bw_\mu\cdot\bdr\,d^d r+\bq_\mu\cdot\bdr \,d^d r=[d(\bm{p}_\mu\,d^d r)/dt]\cdot\bdr$ for any smooth infinitesimal virtual displacement field $\bdr$. The former relation can be integrated on the whole system:
\begin{equation}
\int_{\mathbb{R}^d} \bw_\mu\cdot\bdr\,d^d r+\int_{\mathbb{R}^d} \bq_\mu\cdot\bdr \,d^d r=\int_{\mathbb{R}^d}\frac{d(\bm{p}_\mu\,d^d r)}{dt}\cdot\bdr\,.
\end{equation}
The first integral in Eq.~(\ref{trav_virt}) represents the work $\delta W_\mu$ of the external forces on the whole system, and is therefore equal to the variation $\delta E_\mu$ of the energy of the system during the infinitesimal displacement. Thus, averaging over the microstates $\mu\to([\phi],\bl)$, we obtain
\begin{align}
\langle\delta E_\mu\rangle_{\mu\to([\phi],\bl)}&=-\int_{\mathbb{R}^d} \bq([\phi],\bl)\cdot\bdr \,d^d r\nonumber\\
&+\int_{\mathbb{R}^d}\left\langle\frac{d(\bm{p}_\mu\,d^d r)}{dt}\right\rangle_{\mu\to([\phi],\bl)}\cdot\bdr\,,
\label{trav_virt}
\end{align}
where we have used Eq.~(\ref{cg_q}).
Neglecting the average inertial term in Eq.~(\ref{trav_virt}) yields
\begin{equation}
\langle\delta E_\mu\rangle_{\mu\to([\phi],\bl)}=-\int_{\mathbb{R}^d} \bq([\phi],\bl)\cdot\bdr\,d^d r\,.
\label{varEmu}
\end{equation}
It is possible to neglect the inertial term even out of equilibrium provided that the timescales considered are sufficiently large, as in case of the Langevin equation \cite{Sekimoto}.
 
Let us consider the variation $\delta H$ of the coarse-grained Hamiltonian $H$ during the virtual displacement $\bdr$. Let us denote by $\tilde\phi=\phi+\delta\phi$ the field after the displacement, and by $\bl+\bdl$ the position of the center of mass of the inclusion after the displacement. We can write at first order
\begin{align}
\delta H([\phi],\bl)&\equiv H([\tilde\phi],\bl+\bdl)-H([\phi],\bl)\nonumber\\
&=\frac{k_\mathrm{B}T}{Z([\phi],\bl)}\left(Z([\phi],\bl)-Z([\tilde\phi],\bl+\bdl)\right)\,.
\end{align}
We assume that the smooth infinitesimal virtual displacement $\bdr$ yields a one-to-one mapping of each fundamental microstate $\mu\to([\phi],\bl)$ to a fundamental microstate $\mu'\to([\tilde\phi],\bl+\bdl)$. In the present Appendix, we do not discuss the way $\phi$ is affected by the infinitesimal displacement ---this problem is dealt with in the body of our paper--- but we only assume that $\phi$ is modified by an amount $\delta\phi$, which is a function of $\bdr$. Then, 
\begin{align}
Z([\tilde\phi],\bl+\bdl)&=\sum_{\mu'\to([\tilde\phi],\bl+\bdl)}e^{-\beta E_{\mu'}}\nonumber\\
&=\sum_{\mu\to([\phi],\bl)}e^{-\beta (E_{\mu}+\delta E_\mu)}\,,
\end{align}
and we obtain at first order
\begin{equation}
\delta H([\phi],\bl)=\langle \delta E_\mu \rangle_{\mu\to([\phi],\bl)}\,,
\end{equation}
so Eq.~(\ref{varEmu}) becomes
\begin{equation}
\delta H([\phi],\bl)=-\int_{\mathbb{R}^d} \bq([\phi],\bl)\cdot\bdr\,d^d r\,.
\label{varE}
\end{equation}

Assuming that $\bdr$ is constant in $\ca$, equal to $\bdl$ (so that the inclusion undergoes a translation), Eq.~(\ref{varE}) becomes
\begin{equation}
\delta H([\phi],\bl)= -\bff([\phi],\bl) \cdot\bdl-\int_\cb \bq([\phi],\bl)\cdot\bdr \,d^d r\,,
\label{bilan_incl}
\end{equation}
where 
\begin{equation}
\bff([\phi],\bl)=\int_\ca  \bq([\phi],\bl)\,d^d r
\label{for_dens}
\end{equation}
is the force exerted by the fluid medium on the inclusion in the coarse-grained description. Thus, Eq.~(\ref{bilan_incl}) gives
\begin{equation}
\bff([\phi],\bl)=-\frac{\partial H([\phi],\bl)}{\partial\ell_i}\be_i\,,
\end{equation}
which corresponds to the fundamental definition Eq.~(\ref{def_force}).

Note that the force $\bff([\phi],\bl)$ is relevant at the coarse-graining level where the system is described by $([\phi],\bl)$. It is a force averaged over the microstates $\mu\to([\phi],\bl)$: indeed, in the fundamental microstate $\mu$, the force exerted by the medium on the inclusion is 
\begin{equation}
\bff_\mu=\int_\ca  \bq_\mu\,d^d r\,,
\label{fmu}
\end{equation}
which verifies
\begin{equation}
\langle\bff_\mu\rangle_{\mu\to([\phi],\bl)}=\int_\ca  \langle\bq_\mu\rangle_{\mu\to([\phi],\bl)}\,d^d r=\bff([\phi],\bl)\,,
\label{fmu_fphi}
\end{equation}
where we have used Eq.~(\ref{cg_q}) and~(\ref{for_dens}).

If there are singularities in the energy density of the system at the boundaries of $\ca$, the mathematical definition $\bff=\int_\ca  \bq\,d^d r$ can become ambiguous. This is the case for instance when such singularities yield Dirac terms in $\bq$ at the boundaries of $\ca$ (see, e.g., Appendix~\ref{Ttilde_der}). However, the force $\bff$ must take into account all the terms that come from the presence of the inclusion, including boundary terms. Thus, in general, the integral over $\ca$ has to be carried out using the following procedure:
\begin{equation}
\int_\ca  \bq\,d^d r=\lim_{\epsilon\to 0}\int_{\ca_\epsilon} \bq\,d^d r\,,
\label{procedure}
\end{equation}
where $\ca_\epsilon$ contains $\ca$ and its boundary $\cs$ plus a shell of hypervolume $\epsilon$, so that each point of $\ca$ and $\cs$ is interior to $\ca_\epsilon$ for all $\epsilon>0$. This procedure amounts to performing the integral infinitesimally outside the inclusion, in order to ensure that the whole inclusion is enclosed in our hypersurface of integration.
Using this procedure does not change the result in the case where $\bq$ is a piecewise continuous function. 

Note that we can also extract the hypervolume density $\bq$ of internal forces in the medium, at the coarse-graining level where the system is described by $([\phi],\bl)$, from Eq.~(\ref{varE}): 
\begin{equation}
\bq([\phi],\bl)=-\frac{\delta H([\phi],\bl)}{\delta r_i}\be_i.
\end{equation}

\section{Derivation of the stress tensor $\bT$ of the fluid medium}
\label{T_der}
In this Appendix, we consider a fluid medium with Hamiltonian density $\ch(\phi,\bm{\nabla}\phi)$ without any inclusion. The stress tensor associated to $\ch$ can be derived from Noether's theorem (see, e.g., Ref.~\cite{Burgess}). Indeed, it is the Noether current associated to the translation invariance of the system. It is obtained by considering the following infinitesimal translation:
\begin{equation}
\left\{ \begin{array}{lll}
\br & \rightarrow & \br+\bdr\,,\\
\phi(\br) & \rightarrow & \phi(\br)+\delta\phi(\br)\,.
\end{array} \right.
\end{equation}
with constant $\bdr$, and with $\delta \phi(\br)=-\bm{\nabla}\phi(\br)\cdot\bdr(\br)$, i.e., $\delta^T\!\phi\equiv0$.
The stress tensor given directly by Noether's theorem is often called the \emph{canonical} stress tensor (or more generally, in time-dependent field theories, the canonical energy-impulsion tensor). Indeed, it is its divergence that appears in the conservation law associated with translation invariance, so any tensor with vanishing divergence can be freely added to this canonical stress tensor. Such modifications of the stress tensor are often used in field theory, for instance to ensure its symmetry or its scale or conformal invariance \cite{DiFrancesco}. However, in the case of perturbative embedded inclusions, where the stress tensor is defined everywhere in the system, even inside the inclusions (see Appendix~\ref{Ttilde_der}), Stokes' theorem can be applied to the integral of the stress tensor on any closed hypersurface, so that the divergence-free ``improvement'' terms do not contribute to Casimir-like forces (this argument is used in Ref.~\cite{Kondrat09}). We will not discuss these terms further.

In the present work, our field-theoretic model describes a fluid medium where the interactions are supposed to be short-ranged. Thus, another way of constructing the stress tensor of the medium is to adapt the definition of continuum mechanics. Let us define the stress tensor $\bT$ by the relation
\begin{equation}
df_i=T_{ij}n_j \, d^{d-1} r\,,
\end{equation}
where $\bm{df}$ is the infinitesimal force that one side of the medium (side $1$) exerts onto the other side (side $2$) through the hypersurface $d^{d-1} r$, and $\bm{n}$ denotes the normal to this hypersurface directed toward side $1$ \cite{Gurtin, Batchelor}. The stress tensor $\bT$ of the medium can be determined from its energy $H$ using the principle of virtual work.
For this, we start by cutting (virtually) a subpart $\ca$ of the medium, with energy 
\begin{equation}
H_\ca=\int_\ca\ch(\phi,\bm{\nabla}\phi)\,d^d r\,.
\label{defT}
\end{equation}
Let us call $\cb$ the rest of the medium (i.e., $\cb=\mathbb{R}^d\smallsetminus\ca$), and $\cs$ the interface between $\ca$ and $\cb$.
Let us now consider an infinitesimal transformation of the medium, as defined in Eq.~(\ref{trf_inf}). We consider that each fluid particle keeps the same $\phi$ during the displacement: $\delta \phi(\br)=-\bm{\nabla}\phi(\br)\cdot\bdr(\br)$, i.e., $\delta^T\!\phi\equiv0$. The variation during this transformation of the energy $H_\ca$ of the closed system initially in $\ca$ reads, at first order in $\epsilon$:
\begin{align}
\delta H_\ca=&-\int_\ca\left[\frac{\partial\ch}{\partial\phi}-\partial_j\left(\frac{\partial\ch}{\partial(\partial_j\phi)}\right)\right] \partial_i\phi\,\delta r_i\,d^d r\nonumber\\
&+\int_\cs\left[\ch\delta_{ij}-\frac{\partial\ch}{\partial(\partial_j\phi)}\partial_i\phi\right]n_j\,\delta r_i\,d^{d-1} r\,.
\end{align}
This variation of energy can be equated to the work $\delta W$ done during the transformation by the external forces acting on the closed system initially in $\ca$:
\begin{equation}
\delta W=\int_\ca w_i\,\delta r_i \,d^d r+\int_\cs T_{ij}n_j\,\delta r_i\,d^{d-1}r\,,
\end{equation}
where $\bw$ is the hypersurface density of forces exerted by the exterior on the fluid medium, so that the integral on $\ca$ represents the forces exerted by the exterior of the medium on the closed system initially in $\ca$. Meanwhile, the integral on $\cs$ represents the force exerted by the rest of the medium on the closed system initially in $\ca$, which has been expressed using the stress tensor thanks to its definition Eq.~(\ref{defT}).
Since the energy balance $\delta H=\delta W$ must hold for any virtual deformation field $\bdr$, we can identify the stress tensor of the medium:
\begin{equation}
T_{ij}=\ch\delta_{ij}-\frac{\partial\ch}{\partial(\partial_j\phi)}\partial_i\phi\,.
\end{equation}
In addition, the identification of the bulk term gives the hypervolume density $\bq$ of internal forces in the medium. Neglecting inertia, which is possible if the timescales considered are sufficiently large, Newton's second law applied to each fluid particle of the medium gives $\bq=-\bw$, where $\bw$ represents the hypervolume density of external forces (see Appendix~\ref{sec_def}). Thus, we obtain:
\begin{equation}
q_i=-w_i=\left[\frac{\partial\ch}{\partial\phi}-\partial_j\left(\frac{\partial\ch}{\partial(\partial_j\phi)}\right)\right] \partial_i\phi=\partial_j T_{ij}\,.
\end{equation}

This derivation of the stress tensor shows that the stress tensor itself is built \emph{by assuming that each fluid particle keeps the same $\phi$ during a displacement}. It is therefore not surprising that we find the stress tensor when we calculate the force $\bff^{(2)}$ under this assumption.

With our mechanical definition of the stress tensor, we have obtained, from the principle of virtual work, a stress tensor $\bT$ that is identical to the canonical stress tensor $\bT^c$ given by Noether's theorem. In field theory, the stress tensor $\bT^c$ generally written is in fact $T^c_{ij}=-T_{ji}$ \cite{Burgess, DiFrancesco}, but this apparent difference is just a matter of convention. Note that our stress tensor $\bT$ is fully defined (not up to a term with vanishing divergence) because we have asked it to give the force exchanged through any infinitesimal hypersurface.

\section{Stress tensor $\bT'$ of the fluid medium with a perturbative embedded inclusion}
\label{Ttilde_der}
Let us now consider the fluid medium with a perturbative embedded inclusion in it: its effective Hamiltonian $H$ corresponds to Eq.~(\ref{E_transp}). Carrying out the same reasoning as in the previous section, using a generic infinitesimal deformation $\bdr$, enables to identify the stress tensor $\bm{T}'$ of the composite medium comprising the perturbative inclusion:
\begin{equation}
T'_{ij}=\left(\ch+V \,\bm{1}_\ca\right)\delta_{ij}-\frac{\partial\ch}{\partial(\partial_j\phi)}\partial_i\phi\,.
\end{equation}

The same reasoning also enables to identify the hypervolume density of internal forces at each point of the composite medium as $ q'_i=\partial_j T'_{ij}$. Explicitly, it gives
\begin{equation}
\bq'=\frac{\delta H}{\delta\phi}\bm{\nabla}\phi+V\,\bm{\nabla}\bm{1}_\ca\,,
\label{qp}
\end{equation}
where $H$ is defined in Eq.~(\ref{E_transp}), and its functional derivative with respect to $\phi$ is given by Eq.~(\ref{derfct}).
Note that the gradient of $\bm{1}_\ca$, and thus $\bq'$, features Dirac singularities on the contour of $\ca$, i.e., on $\cs$.

\section{Pointlike embedded inclusion or non-embedded influencing object}
\label{pointlike}
Let us consider a pointlike object in $\br=\bl$ that is coupled to $\phi(\bl)$:
\begin{align}
H&=\int_{\mathbb{R}^d}\ch(\phi,\bm{\nabla}\phi)\,d^d r+V(\phi(\bl))\nonumber\\
&=\int_{\mathbb{R}^d}\left[\ch+V\delta(\br-\bl)\right]\,d^d r\,.
\end{align}
Here, the total variation of $H$ during the infinitesimal transformation (\ref{trf_inf}) reads, at first order:
\begin{align}
\delta H=&\int_{\mathbb{R}^d}\left[\frac{\partial\ch}{\partial\phi}-\partial_i \left(\frac{\partial\ch}{\partial(\partial_i\phi)} \right)\right]\,\delta\phi\,d^d r \nonumber\\
&+\frac{\partial V}{\partial \phi}(\phi(\bl))\left[\delta\phi+\bm{\nabla}\phi\cdot \bdl\right]\,.
\label{vartot_pct}
\end{align}

Thus, the first route, which corresponds to $\delta\phi\equiv 0$, yields
\begin{equation}
\bff^{(1)}=-\bm{\nabla}\phi(\bl)\,\frac{\partial V}{\partial \phi}(\phi(\bl))\,.
\label{f1_pct}
\end{equation}

Meanwhile, if we follow the second route, i.e., if $\delta^\mathrm{T}\!\phi=\delta\phi+\bm{\nabla}\phi\cdot\bdr\equiv 0$, Eq.~(\ref{vartot_pct}) becomes
\begin{align}
\delta H&=-\int_{\mathbb{R}^d}\left[\frac{\partial\ch}{\partial\phi}-\partial_j \left(\frac{\partial\ch}{\partial(\partial_j\phi)} \right)\right] \partial_i\phi\,\delta r_i\,d^d r\nonumber\\
&=-\int_{\mathbb{R}^d}\partial_j T_{ij}\,\delta r_i\,d^d r\,,
\end{align}
where we have used Eq.~(\ref{divT}).
In spite of this simple expression, we must remember that the energy density $\ch+V\delta(\br-\bl)$ has a singularity in $\bl$. To deal with it, let us write 
\begin{align}
\delta H=&-\lim_{\epsilon\to 0}\int_{ \mathbb{R}^d\smallsetminus B_\epsilon^{\bl} } \partial_j T_{ij}\,\delta r_i\,d^d r\nonumber\\
&-\left\{\lim_{\epsilon\to 0}\int_{B_\epsilon^{\bl}}\partial_j T_{ij}\,d^d r\right\}\delta \ell_i\,.
\end{align}
where $B_\epsilon^{\bl}$ denotes the hyperball of radius $\epsilon$ centered on $\bl$.
This relation enables to identify $\bff^{(2)}$ as
\begin{align}
\bff^{(2)}&=\lim_{\epsilon\to 0}\int_{B_\epsilon^{\bl}}\partial_j T_{ij}\,d^d r\,\be_i\nonumber\\
&=\lim_{\epsilon\to 0}\int_{S_\epsilon^{\bl}}T_{ij} n_j \,d^{d-1} r\,\be_i\,,
\label{f2_pct}
\end{align}
where $S_\epsilon^{\bl}$ denotes the hypersphere of radius $\epsilon$ centered on $\bl$ and $\bm{n}$ is its exterior normal.

Thus, the difference between the two ways of keeping $\phi$ constant while varying $\bl$ remains for pointlike objects. As in the case of extended objects, the force $\bff^{(2)}$ is adapted to a pointlike embedded inclusion, while the force $\bff^{(1)}$ is adapted to a non-embedded pointlike influencing object.

\bibliographystyle{apsrev4-1}

\end{document}